# Optimizing Next Generation Wireless BAN with Prioritized Access for Heterogeneous Traffic

Shama Siddiqui*
*DHA Suffa University, Pakistan.*
*South East Technological University, Ireland*
shamasid@hotmail.com
0000-0002-3547-0307

Indrakshi Dey
*South East Technological University*
*Waterford, Ireland*
Indrakshi.Dey@WaltonInstitute.ie
0000-0001-9669-6417

***Abstract*—Efficient management of heterogeneous traffic with varying priorities is critical in Wireless Body Area Networks (WBANs). The priority mechanisms embedded in Media Access Control (MAC) schemes largely govern the performance of WBAN in terms of reliability, delay and energy efficiency. Minimizing the delay between packet generation and reception is critical for enhancing WBAN performance and associated health outcomes; however, delay optimization must be tailored to each traffic priority. In this work, we proposed a novel priority-based MAC protocol, Adaptive and Dynamic Polling MAC for Prioritized Traffic (ADP$^2$-MAC), designed to support heterogeneous traffic in WBANs. The protocol utilizes a probabilistic approach to dynamically determine channel polling/listening intervals. ADP$^2$-MAC not only identifies traffic arrival patterns to determine optimal polling intervals but also interrupts the transmission of lower-priority data when urgent packets are expected. The performance of ADP$^2$-MAC has been compared with the MAC protocol for Variable Data Rates (MVDR) which supports heterogeneous traffic by assigning different data rates based on traffic priority. ADP$^2$-MAC outperforms MVDR due to its use of probabilistic polling intervals and an interruption mechanism designed to efficiently handle urgent-priority data.**

## I. Introduction

Efficient operation of WBANs requires guaranteed delivery of critical priority packets in a timely manner. The rapid growth of Internet of Medical Things (IoMT) has introduced significant challenges in managing wireless network resources across heterogeneous devices, applications, and traffic types [1]. In a typical healthcare scenario, there are multiple traffic priorities, such as emergency or critical, periodic or normal, and on-demand. For example, continuous monitoring of vital signs such as heart rate and blood pressure generates periodic traffic, while sudden alerts indicating abnormal conditions represent emergency traffic [2]. In the absence of efficient MAC protocols, critical traffic can suffer from excessive delays and elevated energy consumption caused by limited access to required channel resources [3]. Therefore, implementing traffic-aware MAC protocols is essential to ensure prompt and reliable transmission of high-priority medical data, thereby maintaining quality of service in healthcare applications.

Various approaches have been employed in priority-based MAC protocols for WBAN. A multichannel hybrid MAC, (MC-HYMAC) has been proposed in [4] that combines the benefits of Carrier Sense Multiple Access with Collision Avoidance (CSMA/CA) and Time Division Multiple Access (TDMA) protocols. Adaptive MAC (ADT-MAC) [5] proposed the novel concept of using dynamic super-frame structure in IEEE 802.15.6 standard to accommodate periodic and emergency traffic of WBAN. Moreover, The MAC protocol for Variable Data Rates (MVDR) provides differentiated priority by assigning different data rates to traffic based on its priority level [6]: the critical traffic is assigned a higher data rate, representing a better service in terms of delay, energy consumption, and packet delivery ratio.

In this work, we propose ADP$^2$-MAC, which is an asynchronous Adaptive and Dynamic MAC Protocol for Prioritized Traffic; the protocol is particularly well-suited for WBAN applications due to its ability to serve traffic with varying priority levels. ADP$^2$-MAC has been proposed as an extension to a previous protocol, Adaptive and Dynamic Polling-MAC (ADP-MAC) [7]; ADP-MAC introduced the novel concept of using polling interval distributions for efficient management of synchronization between packet arrival and node's wake-up instant. Co-efficient of variation (Cv) of the incoming traffic patterns was analyzed in order to predict the wake-up intervals of the nodes, leading to energy and delay efficient operation. The dynamic nature of ADP-MAC makes it well-adapted to accommodate diverse traffic patterns such as Constant Bit Rate (CBR), Poisson arrivals, and bursty transmissions [8]. We further detail the operation of ADP-MAC in section II.

To the best of our knowledge, no previous MAC protocol has proposed a prioritization method using dynamic polling interval distributions. The main contributions of this paper are listed below: leftmargin=*

- To propose an asynchronous Adaptive and Dynamic MAC Protocol, ADP$^2$-MAC for Prioritized Traffic, specifically required in WBAN.
- To study the performance of ADP$^2$-MAC for heterogeneous traffic in WBAN.
- To perform a comparative analysis of ADP$^2$-MAC against MVDR, which differentiates traffic priorities by allocating distinct data rates.

The rest of this paper has been organized as follows: section

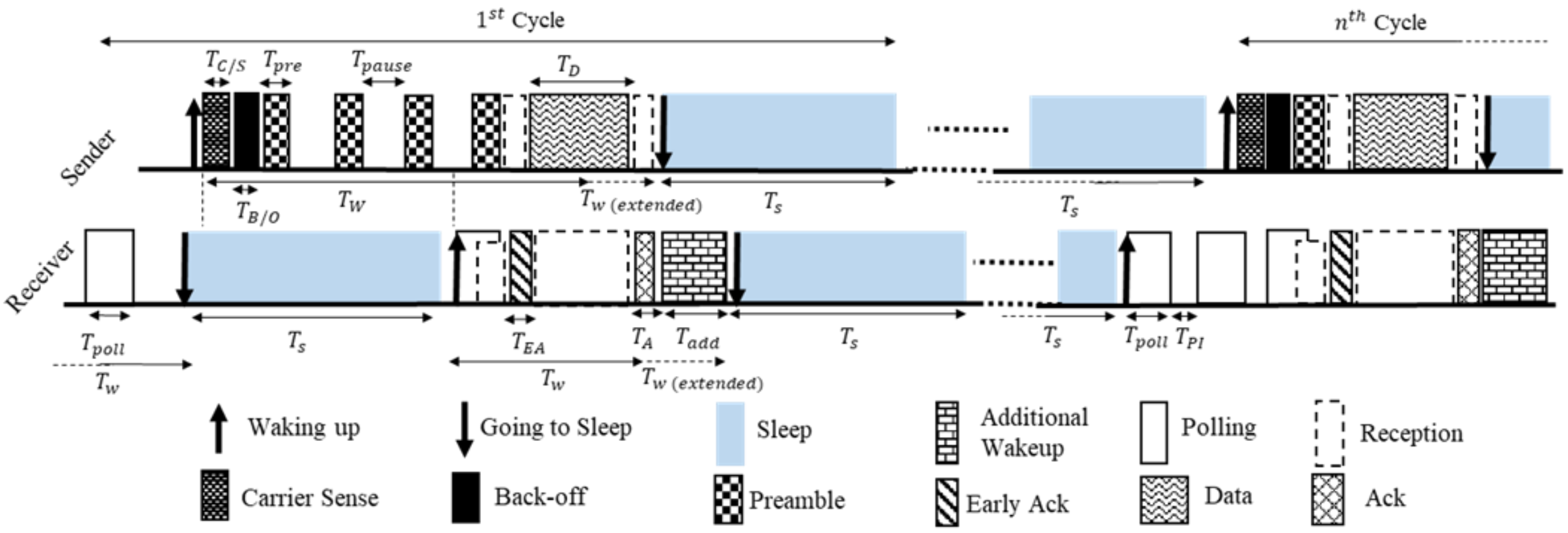

Fig. 1: Basic Operation of ADP-MAC.

II presents a brief literature review; section III details the proposed design of ADP²-MAC; section IV describes the experimental setup; section V explains the results and finally, section VI summarizes the work and recommends directions for future research.

## II. Related work

WBANs present several challenges that must be addressed to ensure reliable performance and satisfactory health outcomes. First, it is crucial to assign priorities to data of different types; for example, low-priority periodic data may be generated for routine health monitoring, while high-priority data corresponds to critical events such as falls, cardiac arrests, or accidents [9]. Therefore, MAC protocols designed for WBAN adopt various strategies to guarantee a prioritized channel access to the nodes [10]. In this section, we briefly review some of these strategies and present the ADP-MAC and MVDR protocols in greater detail, as they form the foundation of the present study.

Using multi-channels has been a common approach for dealing with heterogeneous traffic in WBAN. In such protocols, data of different priorities is assigned different channels and nodes are scheduled to listen to the urgent priority channels more frequently [11]. Although this approach ensures that nodes with higher priority data get a quicker and guaranteed channel access [12], there are hardware limitations as not all WBAN nodes support multi-channels [13]. Similarly, use of adaptive Contention Window (CW) makes a conventional priority scheme for WBANs [14]–[16], like other wireless networks; however, relying solely on the CW scheme may lead to starvation for nodes assigned higher CW values.

In [7], ADP-MAC was developed to reduce energy consumption and latency in WSN applications with dynamic traffic arrival patterns. A key innovation of ADP-MAC is its adaptive selection of polling interval distributions based on incoming traffic patterns. By monitoring the Cv of traffic arrivals, the protocol distinguishes between CBR and Poisson arrival patterns. Deterministic polling intervals are used for CBR traffic, while exponential polling intervals are applied for Poisson traffic; this dynamic adjustment ensures optimal performance across varying traffic conditions. Additionally, ADP-MAC incorporates features such as packet concatenation and block acknowledgments, further enhancing energy efficiency by minimizing redundant overhead bytes.

The fundamental operation of ADP-MAC is depicted in Figure 1. In the first cycle, the source node performs carrier sensing and back-off before transmitting preamble strobes, while the receiver remains in a sleep state. Upon waking up and detecting the preamble, the receiver responds by sending an Early Acknowledgment (EA); this concept of using short preambles and EA has been conventionally used in X-MAC [17] to reduce the preamble transmission energy, and delay incurred due to listening to the full preamble even if the receiver is awake. Once the channel access is established, the data packet is transmitted, followed by an acknowledgment from the receiver. Instead of immediately returning to sleep, the receiver remains active for an additional duration, $T_{add}$. If no further packet is received during this period, the receiver transitions into a predefined sleep state. In the n$^{t}h$ cycle, the receiver wakes up and initiates polling, but no activity is detected on the channel. Subsequently, the sender wakes up and begins transmitting preambles, after which the communication proceeds similarly to the first cycle. Further details about operation and implementation of ADP-MAC can be found in [7].

In this work, we used a previous protocol MVDR for comparison with ADP²-MAC. MVDR prioritizes data based on its criticality, assigning higher data rates to vital information to ensure prompt delivery. By incorporating the Guaranteed Time Slots (GTS) in the IEEE 802.15.4 standard super-frame, MVDR minimizes energy consumption and delays associated with channel access. Additionally, it employs node similarity predictions to enhance the reliability of data transmission, particularly for high-priority information. Although MVDR performs well in terms of energy consumption, delay, and packet delivery ratio, it still experiences increased delays and energy usage due to synchronization overhead and the waiting time imposed by the conventional super-frame structure.

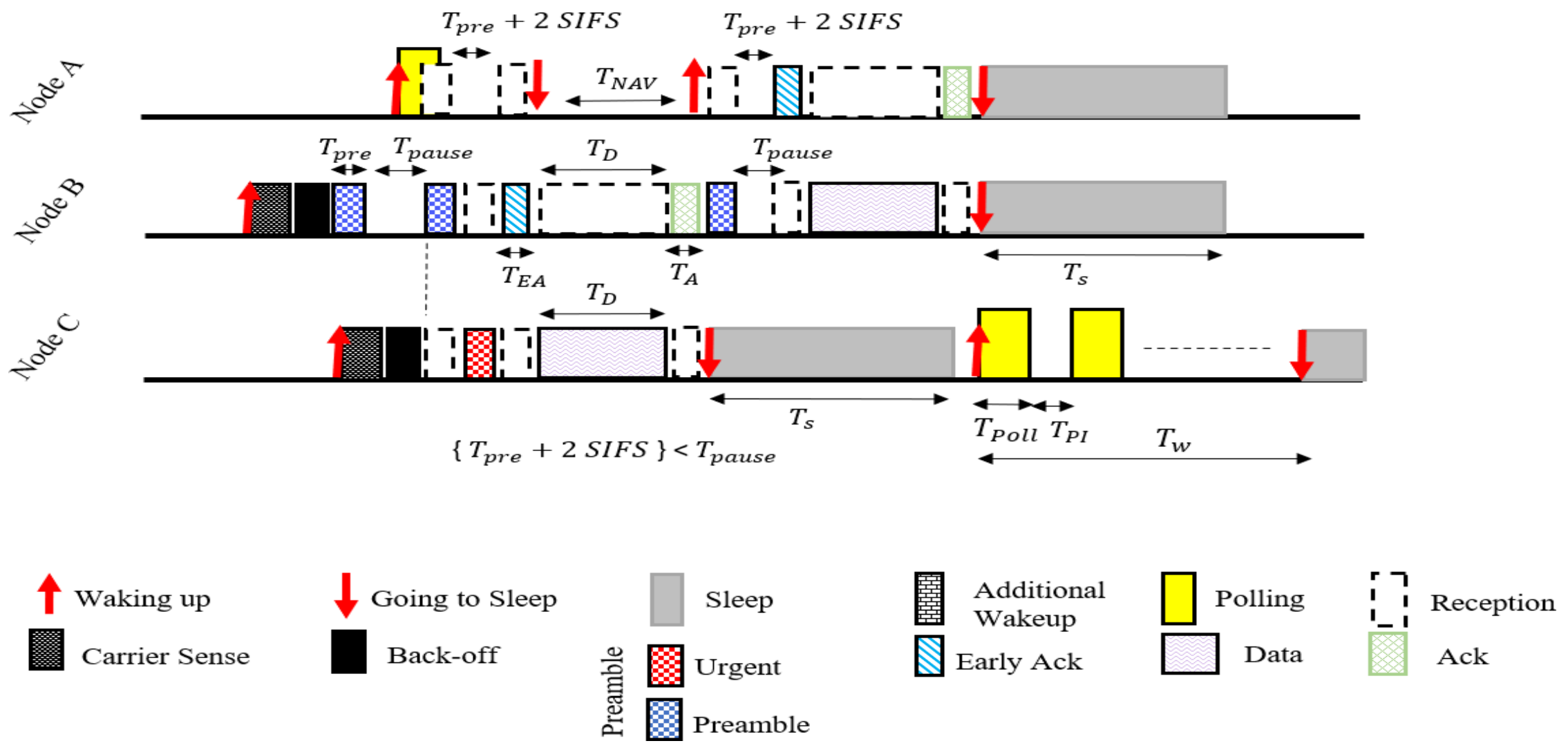

Fig. 2: Timing Diagram Illustrating the Prioritized Transmission of ADP²-MAC.

To summarize, existing MAC protocols offer valuable solutions but fall short in addressing urgent traffic handling without compromising energy efficiency. These gaps motivate the design of ADP²-MAC, which introduces priority-aware interruption and adaptive polling to better serve heterogeneous WBAN traffic.

## III. Protocol Design

In this study, a novel protocol named ADP²-MAC is proposed, building on the existing ADP-MAC [7] framework with enhancements tailored for heterogeneous traffic environments. While ADP-MAC adjusts listening intervals based on traffic arrival patterns, it does not account for traffic priorities, which is a crucial aspect for next-generation WBANs. To meet the stringent requirements of WBANs, it is essential to ensure low latency for time-sensitive physiological data, optimize energy efficiency for battery-operated biosensors, and maintain high reliability for critical health monitoring communications [18]. Therefore, ADP²-MAC introduces priority handling mechanisms, including adaptive CW and transmission interruption strategies. Firstly, the adaptive CW mechanism allows nodes with higher-priority traffic to select smaller CW sizes, enabling shorter waiting periods before transmission initiation [19]. This approach aligns with common MAC strategies for prioritization, where higher-priority nodes reduce contention delays to expedite data transmission. Secondly, we introduce a mechanism to detect ongoing urgent data requests on the channel; if a node attempting to transmit a normal-priority packet identifies that another node needs to send urgent data, it relinquishes its transmission opportunity to prioritize the urgent packet.

We illustrate the concepts of dynamic polling intervals integrated with prioritized interruption proposed in ADP²-MAC, in figure 2. It is important to note that the short preamble bytes are designed to include information about the priority of data packets. In figure 2, it is assumed that nodes A and C are hidden for each other. Node B started the transmission operation for a normal packet destined for node A; B is shown to perform the carrier sense, followed by back-off and first preamble strobe transmission. Node A polls the channel and receives the preamble indicating a transmission of normal packet from node B. However, it does not immediately respond with EA; rather it waits for a period, $T_{wait} = T_{pre} + 2 * SIFS$ to allow interruption to some node which might have an urgent packet. $T_{wait}$ has been designed such that a node having urgent data can efficiently interrupt the ongoing transmission activity for low priority packets; to ensure this, we maintain the relation $(T_{pre} + 2 * SIFS) < T_{pause}$. Thus, the duration of $T_{wait}$ implies that a preamble associated with urgent packet will be transmitted before the node with normal packet could send its next preamble strobe.

Meanwhile, node C had started the transmission operation for an urgent packet destined for node B. Although node A could not listen to the transmission of C, it could still receive an EA from node B. Thus, it gets to know that node B is now engaged in a reception of urgent packet from some node; A, therefore, sets a NAV timer and goes to sleep; it again polls the channel once the reception activity of B is expected to finish. It is to be noted that if a preamble is received indicating an urgent packet, the receiver node would not wait, but will transmit an EA immediately. Node C sends its urgent packet, receives an acknowledgment and goes to sleep. Once the NAV expires and A begins polling again, node B could not now send its pending normal packet. Both nodes A and B will now sleep for the pre-defined sleep interval.

## IV. Experimental Settings

We conducted simulations using the Avrora simulator, targeting mica2 motes due to their compatibility with low-power wireless applications. ADP²-MAC was implemented by extending Avrora's radio and event-handling modules to support traffic prioritization, dynamic queue management, and adaptive duty cycling. To simulate heterogeneous traffic in a WBAN context, we generated two types of data: urgent traffic, modeled as Poisson arrivals ($\lambda = X$) to represent unpredictable, time-critical events such as falls or abnormal heart activity, and normal traffic, modeled as constant bit rate (CBR) to reflect periodic monitoring tasks like temperature or pulse readings. The network topology consisted of 8 nodes arranged linearly, where each node was capable of both sensing and relaying packets, enabling multi-hop communication. This reflects realistic WBAN deployments where multiple body-worn sensors may detect the same critical event and forward data toward a central sink. Static forwarding rules were used to relay packets upstream. Each simulation ran until 1,000 packets were successfully delivered across the network.

TABLE I: Simulation Parameters

| Simulation Parameters | Value |
|---|---|
| Bit rate | 18.78 kbps |
| Simulation Duration | Until 1000 data packets transmitted |
| Arrival Patterns | CBR (Normal)<br>Poisson (Urgent) |
| Polling Interval Distributions | Deterministic / Exponential / Dynamic |
| Total Nodes | 8 |
| Mean Message Generation Interval | Variable |
| Duration of Each Cycle $T_{cycle}$ | 10 sec |
| Size of Contention Window | Up to 32 Slots.<br>Variable for Urgent traffic |
| Wake-up Time $T_W$ | 300 msec |
| Sleep Time $T_S$ | 9700 msec |
| Polling Duration $T_{poll}$ | 20 msec |
| Polling Interval $T_{PI}$ | 50 msec |
| Time Required to Transmit Each Preamble Strobe $T_{pre}$ | 10 msec |
| Pause between 2 Preamble Strobes | 10 msec |
| Polling Duration | 20 msec |
| Additional Wakeup Time $T_{add}$ | 100 msec |
| Time required to Transmit Early ACK $T_{EA}$ | 10 msec |
| Time required to Transmit Data $T_D$ | 25 msec |
| Time required to Transmit ACK/Block ACK | 10.05 msec |
| Propagation Delay Time required to detect the channel activity | 7 msec |

Energy consumption was evaluated using Avrora's built-in energy model, which accounts for hardware-level operations including radio transmission, reception, CPU activity, and idle listening. We recorded both per-node and network-wide cumulative energy consumption to assess protocol efficiency. Detailed simulation parameters, such as transmission power, data rates, and queue sizes, are provided in Table 1, with reference configurations aligned with prior work [7] for consistency.

## V. Results and Discussions

The energy efficiency and delay performance of ADP²-MAC were evaluated by varying the message generation intervals for both urgent and normal data. The protocol's performance was benchmarked against ADP-MAC and MVDR using identical simulation settings; the results have been shown in figure 3.

Figure 3-a shows that as the message generation interval increases, the energy consumption of all protocols reduces. This is becasue as the interval between message generations increases, the frequency of collisions reduces, leading to fewer retransmissions and reduced overall channel contention. Consequently, nodes spend less time in active communication states, resulting in lower power consumption. The energy consumption of ADP²-MAC for urgent traffic has been found to be the lowest, and the highest trend has been observed for MVDR when serving normal traffic. In fact, for each protocol, the energy consumed for urgent packets is lower as compared to the normal traffic. MVDR demonstrated significantly lower performance compared to both ADP-MAC and ADP²-MAC, as it is based on 802.15.4 super-frame structure, leading to synchronization overhead. Moreover, MVDR does not offer a dynamic scheduling of channel listening periods or prioritized interruption; this implies that once the channel is assigned to

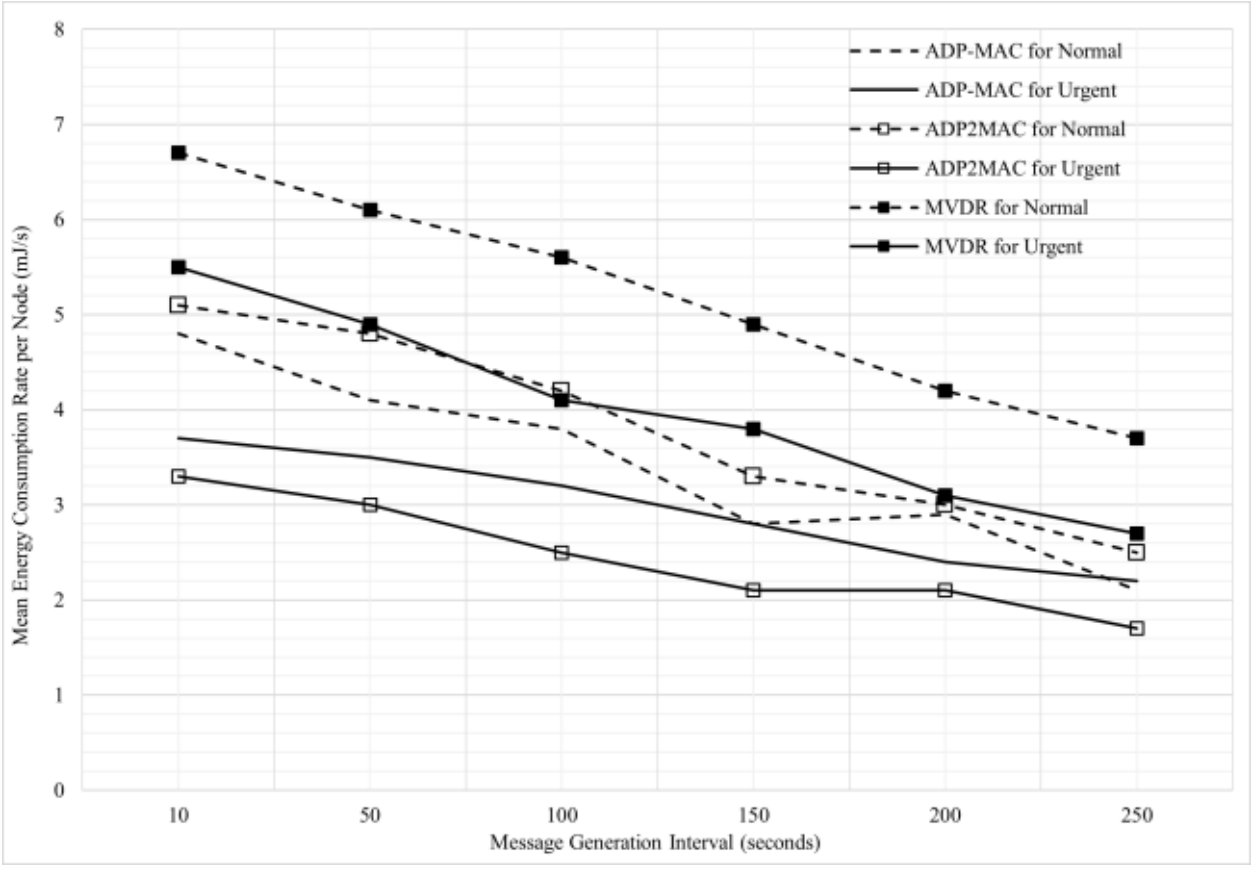

(a)

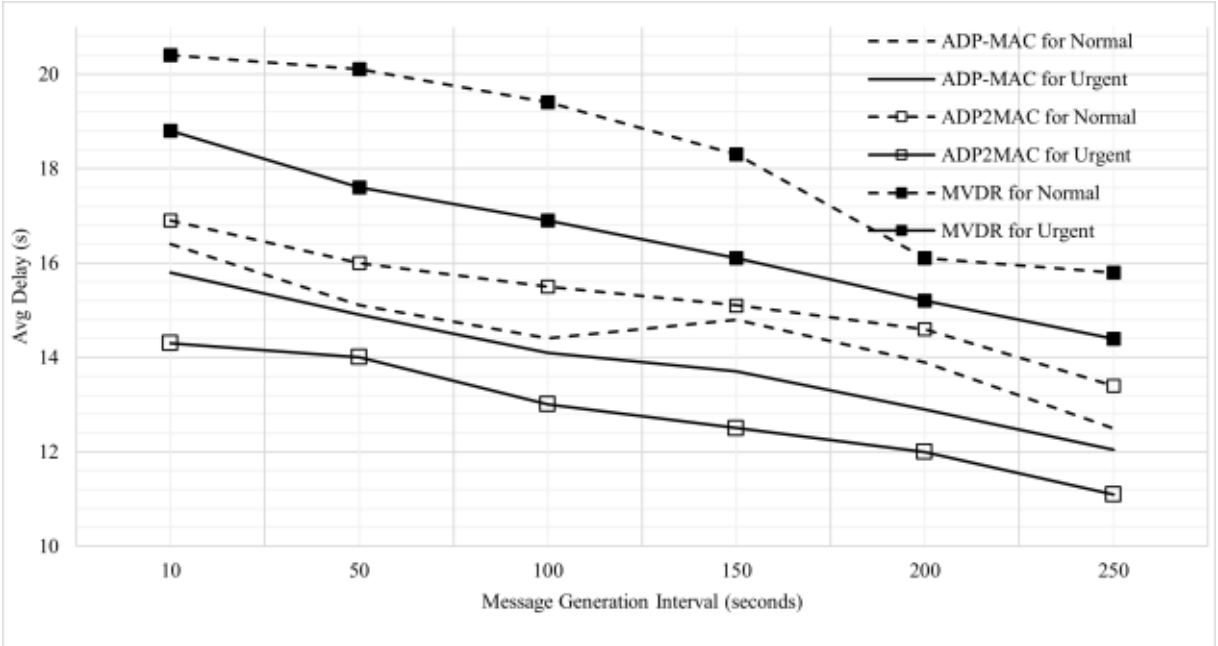

(b)

Fig. 3: Simulation Results: (a) Energy Consumption, (b) Average Delay.

the normal priority traffic during the Contention Access Period (CAP), the nodes with urgent traffic will need to wait for the next super-frame. All these factors contribute towards a higher energy consumption of MVDR.

Moreover, Figure 3-b illustrates the comparative evaluation of three protocols in terms of average delay. The trend for delay is also decreasing with the message generation interval due to lower contention, collisions and retransmissions. Just in correspondence with figure 3-a, delay also exhibits same trends in figure 3-b: the delay for MVDR with normal traffic is the highest, whereas that for ADP²-MAC is found to be the lowest. Although MVDR uses different data rates for high priority traffic, the delay remains significantly higher than ADP-MAC and ADP²-MAC because of the synchronization overhead and non-optimal usage of channel resources. It is important to note that ADP²-MAC reduces the delay for urgent traffic mainly due to letting the nodes with urgent traffic transmit by interrupting the ongoing transmission activity of normal packets. This feature was not available in either ADP-MAC or MVDR.

## VI. Conclusion and Future Work

This paper presented a novel MAC protocol for WBAN, called ADP²-MAC, which incorporates adaptive and probabilistic polling interval distribution, dynamic contention window adjustment, and a transmission interruption mechanism. These strategies collectively enable the protocol to effectively support heterogeneous traffic types, offering timely channel access for urgent data while conserving energy during routine transmissions. Simulation results using the Avrora emulator demonstrated that ADP²-MAC outperforms both ADP-MAC and MVDR in terms of energy efficiency and packet delay, primarily due to its ability to adapt channel polling and prioritize urgent transmissions.

Future research will focus on implementing and evaluating ADP²-MAC on body sensor prototypes to assess its performance under realistic conditions, including human motion, wireless interference, and physiological signal variability. Such experimental validation will help demonstrate the protocol's robustness and suitability for practical healthcare monitoring. Additionally, exploration of cross-layer optimization techniques will be pursued, integrating information from the application and network layers to dynamically adjust protocol parameters, thereby enhancing responsiveness and energy efficiency for diverse and time-critical WBAN applications.

## VII. Acknowledgment

This contribution is supported by HORIZON-MSCA-2022-SE-01-01 project COALESCE under Grant Number 10113073.